\documentclass[aps,twocolumn]{revtex4}

\usepackage{graphicx}
\usepackage{dcolumn}
\usepackage{bm}

\begin{document}

\title{Algebraic Fermi liquid from phase fluctuations:
``topological" fermions, vortex ``berryons" and
QED$_3$ theory of cuprate superconductors}

\author{M. Franz}
\altaffiliation[Permanent address: ]{Department of Physics and Astronomy,
University of British Columbia, Vancouver, BC, Canada V6T 1Z1}
\author{Z. Te\v{s}anovi\'c}
\altaffiliation[Permanent address: ]{Department of Physics and Astronomy,
Johns Hopkins University, Baltimore, MD 21218}
\affiliation{Institute for Theoretical Physics, University of California,
Santa Barbara, CA 93106
\\
\rm(\today)}

\begin{abstract}
Within the phase fluctuation model for the pseudogap state of
cuprate superconductors we identify a novel statistical
``Berry phase" interaction between the nodal quasiparticles and
fluctuating vortex-antivortex excitations. 
The effective action describing this model
assumes the form of an anisotropic
Euclidean quantum electrodynamics in (2+1) dimensions (QED$_3$) and
naturally generates non-Fermi liquid behavior
for its fermionic excitations. The doping axis in the $x$-$T$
phase diagram emerges as a {\em quantum critical line}
which regulates the low energy fermiology.
\end{abstract}

\maketitle

Perhaps the most intriguing property of high temperature superconductors
is the anomalous character of their normal state \cite{anderson}.
This ``strange metal"
stands in stark contrast to the relatively benign
features of the superconducting
phase which can be understood rather accurately within the
framework of a $d$-wave BCS-like phenomenology
with well defined quasiparticle excitations \cite{millisorenstein}.

In this Letter we propose a theory of the pseudogap
phase in cuprate superconductors based on the following premise:
a successful phenomenology of the strange metal
should be built by starting from a comprehensive understanding of the
adjacent superconducting state and its excitations.
The spirit of our approach is
the traditional one \cite{anderson,pines} but turned upside down.
Usually, the strategy is to first understand the normal state before we
can understand the superconductor.
In the cuprates, however, it is the superconducting
state that appears ``conventional"
and its quasiparticles ``less correlated" and better defined.
Having adopted this ``inverted" paradigm, we proceed to
study the interactions of the quasiparticles with the collective
modes of the system, i.e. fluctuating (anti) vortices (our
strategy here is similar to that of Ref.  \cite{nodalliquid}).
We show that in $d$-wave superconductors these interactions
take a form of a gauge theory which
shares considerable similarity with the quantum electrodynamics
in (2+1)-dimensions (QED$_3$). In the superconducting
state, where vortices are {\em bound}, the gauge fields
of the theory are {\em massive} and the low
energy quasiparticles remain well defined excitations.
This is the mundane Fermi liquid state in our inverted paradigm.
In the normal state, however, as vortices {\em unbind},
our QED$_3$ like theory enters its {\em massless} phase and it
abandons this ``inverted Fermi liquid" protectorate 
in favor of a weakly destabilized Fermi liquid characterized by a power law
singularity in the fermion propagator which we call
{\em algebraic Fermi liquid}.
%
\begin{figure}[t]
\includegraphics[width=8cm]{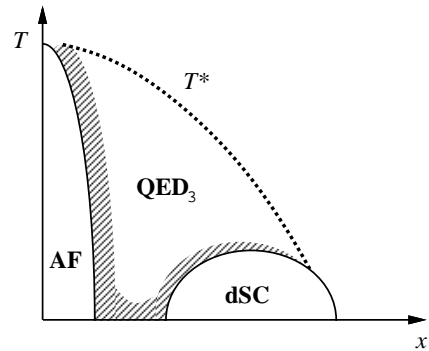}
\caption{\label{fig1}
Phase diagram of a cuprate superconductor.}
\end{figure}
We compute the spectral properties of fermions in our theory and find that
it captures some key qualitative aspects of the available experimental
data.

We concentrate on the portion of the pseudogap
phase above the shaded region
and below $T^*$ in Fig.\ 1. We assume that
Cooper pairs are formed at or somewhat below $T^*$ but the long-range
phase coherence sets in only at the superconducting transition
temperature $T_{sc}\ll$ $T^*$ \cite{emerykivelson}.
Between $T_{sc}$ and  $T^*$ the phase
order is destroyed by unbound vortex-antivortex excitations of the
Cooper pair field \cite{corson,ong,sudbo}. In this pseudogap regime the $d$-wave
superconducting gap is still relatively intact \cite{nodalliquid,emerykivelson}
and the dominant interactions are those of  nodal quasiparticles with
fluctuating vortices. There are {\em two} components of
this interaction: first, vortex fluctuations produce
variations in superfluid velocity which
cause Doppler shift in quasiparticle energies \cite{volovik}.
This effect is classical and
already much studied \cite{franzmillis,dorsey}.
Second, there is a purely quantum ``statistical" interaction,
tied to a geometric ``Berry phase" effect
that winds the phase of a quasiparticle
as it encircles a vortex \cite{ft,marinelli}.
It is this quantum mechanical interaction that ultimately causes the
destruction of the Fermi liquid in the pseudogap phase.

Our starting point is the partition function
\begin{eqnarray}
Z&=&\int{\cal D}\Psi^{\dag}({\bf r},\tau)
\int{\cal D}\Psi({\bf r},\tau)\int{\cal D}\varphi({\bf r},\tau)
\exp{[-S]},\nonumber \\
S&=&\int d\tau\int d^2r\{\Psi^{\dag}\partial_{\tau}
\Psi + \Psi^{\dag}{\cal H}\Psi + (1/g)\Delta^*\Delta \},
\label{action0}
\end{eqnarray}
where $\tau$ is the imaginary time, ${\bf r}=(x,y)$,
$g$ is an effective coupling constant, and
$\Psi^{\dag}=(\bar\psi_{\uparrow},\psi_{\downarrow})$
are the standard Grassmann variables. The Hamiltonian
${\cal H}$ is given by:
\begin{equation}
{\cal H} =\left( \begin{array}{cc}
\hat{\cal H}_e  & \hat{\Delta} \\
\hat{\Delta}^*  & -\hat{\cal H}_e^*
\end{array} \right)
\label{h1}
\end{equation}
with
$\hat{\cal H}_e={1\over 2m}(\hat{\bf p}-{e\over c}{\bf A})^2-\epsilon_F$,
$\hat{\bf p}=-i\nabla$ (we take $\hbar=1$),
and $\hat\Delta$ the $d$-wave pairing operator \cite{ft},
%
%
$\hat{\Delta} =
(1/k_F^{2})\{\hat{p}_x,\{\hat{p}_y,\Delta\}\}
-(i/4k_F^2)\Delta
\bigl (\partial_x\partial_y\varphi\bigr )$,
where $\Delta ({\bf r},\tau) =
|\Delta |\exp(i\varphi ({\bf r},\tau))$ is the center-of-mass
gap function.
$\int{\cal D}\varphi ({\bf r},\tau)$ denotes an integral
over smooth (``spin wave") and singular (vortex) phase
fluctuations. Amplitude fluctuations are suppressed below $T^*$.

It is convenient to eliminate the phase $\varphi ({\bf r},\tau)$
from the pairing term (\ref{h1}) in favor of
$\partial_{\mu}\varphi$ terms [$\mu = (x,y,\tau)$]
in the fermionic action. In order to avoid dealing with
non-single valued wavefunctions we employ
the singular gauge transformation devised in Ref.  \cite{ft}:
\begin{equation}
\bar\psi_{\uparrow}\to\exp{(i\varphi_A)}\bar\psi_{\uparrow},\ \ \
\bar\psi_{\downarrow}\to\exp{(i\varphi_B)}\bar\psi_{\downarrow},
\label{ft}
\end{equation}
where $\varphi_A + \varphi_B
= \varphi$. Here $\varphi_{A(B)}$ is the singular part
of the phase due to $A(B)$ vortex defects: 
$\nabla\times\nabla\varphi_{A(B)}=2\pi\hat{z}\sum_i
q_i\delta({\bf r}-{\bf r}_i^{A(B)})$, with $q_i=\pm 1$ denoting the 
topological charge of $i$-th vortex and 
${\bf r}_i^{A(B)}$ its  position. The labels
$A$ and $B$ represent some convenient but otherwise
arbitrary division of vortex defects (loops or
lines in $\varphi ({\bf r},\tau)$) into two sets.
As discussed in  \cite{ft} this transformation guarantees that the fermionic 
wavefunctions remain 
single-valued and the effect of branch cuts is incorporated directly
into the fermionic part of the action:
$${\cal L}'=\bar\psi_{\uparrow}[\partial_{\tau} +
i(\partial_{\tau}\varphi_A)]\psi_{\uparrow} +
\bar\psi_{\downarrow}[\partial_{\tau} +
i(\partial_{\tau}\varphi_B)]\psi_{\downarrow}
+ \Psi^{\dag}{\cal H}^{\prime}\Psi,
$$
\noindent
where the transformed Hamiltonian ${\cal H}^{\prime}$ is:
\[
\left( \begin{array}{cc}
{1\over 2m}(\hat{\bf \pi}+{\bf v})^2
-\epsilon_F & \hat D \\
\hat D  &
-{1\over 2m}(\hat{\bf \pi}-{\bf v})^2
+\epsilon_F
\end{array} \right),                                                           \]
with
$\hat D =(\Delta_0/2k_F^2)(\hat \pi_x \hat \pi_y+\hat \pi_y \hat \pi_x)$ and
${\hat{\bf \pi}} ={\hat{\bf p}} + {\bf a}$.

The transformation (\ref{ft}) generates a ``Berry gauge potential"
$a_\mu={1\over 2}(\partial_\mu\varphi_A -
\partial_\mu\varphi_B)$
which describes half-flux Aharonov-Bohm scattering of
quasiparticles on vortices and mimics the effect of
branch cuts in quasiparticle-vortex dynamics \cite{ft,marinelli}.
This is in addition to the ``Doppler" gauge field
$v_\mu={1\over 2}(\partial_\mu\varphi_A +
\partial_\mu\varphi_B)$ which denotes the classical part
of the quasiparticle-vortex interaction.
All choices of the sets $A$ and $B$ are
{\em equivalent} -- different choices represent different
singular gauges and $v_\mu$ is
invariant under such transformations.
To symmetrize the partition function with respect
to this singular gauge we define a generalized
transformation (\ref{ft}) as the sum over all possible choices
of $A$ and $B$, i.e., over the entire family of singular
gauge transformations. This is an Ising sum
with $2^{N_l}$ members, where $N_l$
is the total number of
vortex defects in $\varphi ({\bf r},\tau)$.
This symmetrization leads to the new partition function
$Z\to \tilde Z = \int{\cal D}\tilde\Psi^{\dag}
\int{\cal D}\tilde\Psi\int{\cal D}v_\mu\int{\cal D}a_\mu
\exp{[-\int d\tau\int d^2r\tilde{\cal L}]}$
in which the half-flux-to-minus-half-flux (Z$_2$) symmetry of
the singular gauge transformation (\ref{ft}) is manifest:
$$\tilde{\cal L}=\tilde\Psi^{\dag}[(\partial_{\tau}
+ia_\tau)\sigma_0 + iv_\tau\sigma_3]
\tilde\Psi + \tilde\Psi^{\dag}\tilde{\cal H}\tilde\Psi
+{\cal L}_0[v_\mu,a_\mu],$$
\noindent
where ${\cal L}_0$ is the `Jacobian' of the transformation given by
\begin{eqnarray}
e^{-\int d\tau\int d^2r {\cal L}_0}=
2^{-N_l}\sum_{A,B}\int {\cal D}\varphi ({\bf r},\tau) \ \ \ \ \ \ \
\label{action}     \\
\times
\delta[v_\mu - {\textstyle {1\over 2}}(\partial_\mu\varphi_A +
\partial_\mu\varphi_B)]
\delta[a_\mu - {\textstyle {1\over 2}}(\partial_\mu\varphi_A -
\partial_\mu\varphi_B)].\nonumber
\end{eqnarray}
Here $\sigma_\mu$ are the Pauli matrices
and $\tilde{\cal H}={\cal H}^{\prime}$.
We call the quasiparticles $\tilde\Psi^{\dag}=
(\bar{\tilde\psi_{\uparrow}},\tilde\psi_{\downarrow})$
appearing in (\ref{action}) ``topological fermions" (TF's).
TF's are the natural fermionic excitations of the
pseudogapped normal state. They are electrically neutral
and are related to the original quasiparticles by the
inversion of transformation (\ref{ft}).

To proceed we must extract the low energy, long distance properties
of the Jacobian (\ref{action}). This is done by focusing on the
fluctuations of two gauge fields $v_\mu$ and $a_\mu$
in the fluid of vortex excitations.
We use the saddle-point approximation
to compute the leading (quadratic) terms in
${\cal L}_0$ for two cases of interest:
{\em i}) the thermal vortex-antivortex fluctuations in 2D layers and
{\em ii}) the space-time vortex
loop excitations relevant for low temperatures ($T\ll T^*$) in
the underdoped regime (but still above the
shaded region in Fig. 1). The computation is straightforward
but the algebra is laborious and will be presented
elsewhere \cite{qed3long}. Here we quote only the final results whose form is
ultimately dictated by the symmetries of the problem.
For the case {\em i}) 
\begin{equation}
{\cal L}_0 \to {T\over 2\pi^2 n_l}\left[
(\nabla\times{\bf v})^2 +
(\nabla\times{\bf a})^2\right],
\label{2d'}
\end{equation}
where $n_l$ is the average density of free vortex defects.
Both ${\bf v}$ and ${\bf a}$ have a Maxwellian {\em bare} stiffness
and are {\em massless} in the normal
state. As one approaches $T_{sc}$, $n_l\sim \xi_{sc}^{-2}\to 0$,
where $\xi_{sc}(x,T)$ is the superconducting correlation
length, and ${\bf v}$ and ${\bf a}$ become {\em massive}.
Similarly, for the case {\em ii}), the quantum
fluctuations of {\em unbound}
vortex loops result in \cite{qed3long}:
\begin{equation}
{\cal L}_0 \to
{1\over 2\pi^2}\left [ 
K_\tau (\partial\times a)^2_\tau
+\sum_i K_i(\partial\times a)^2_i\right ]
\label{2+1d}
\end{equation}
where $K_\tau, K_i (i=x,y)$ are functions of $x$ and $T$: 
$K_i\sim \xi_{sc}$,
$K_\tau\sim \xi_{sc}^{z}$, $z$ being the dynamical exponent.
The Maxwell form of ${\cal L}_0$ is dictated by symmetry:
the {\em bare} propagators for $v_\mu$ and $a_\mu$,
${\cal D}_v^0({\bf q},i\omega)$ and ${\cal D}_a^0({\bf q},i\omega)$, are
massless in the normal state and massive within a superconductor.
Note that we dropped $v_\mu$ from (\ref{2+1d}) -- the reason for this
is made apparent below.

The physical picture advanced in this Letter
rests on the following observations:
$v_\mu$ couples to the TF ``charge" in the same way as the real
electromagnetic gauge field.
Consequently, if we integrate out TF's in
(\ref{action}) to obtain the renormalized (or dressed) gauge field
propagators ${\cal D}_v({\bf q},i\omega)$ and ${\cal D}_a({\bf q},i\omega)$,
we find that ${\cal D}_v^{-1}({\bf q}\to 0,i\omega=0)\to$ const., i.e. the
Doppler gauge  field $v_\mu$ is {\em massive}.  This is a
consequence of the Meissner response
of TF's. Physically, this means that the integration
over the quasiparticles leads to the familiar
long range interactions between vortices.
In contrast, the ``Berry" gauge field $a_{\mu}$ couples to the TF
{\em spin}. This implies that any contribution of TF's to the stiffness
of $a_{\mu}$ must be {\em massless}:
a singlet superconductor retains the global SU(2) spin symmetry
ensuring that ${\cal D}_a^{-1}({\bf q}= 0,i\omega=0)=0$.

When we combine this with the bare propagators implied by Eqs.\
(\ref{2d'},\ref{2+1d}) the following physical picture emerges.
In the superconducting state, both
$v_\mu$ and $a_{\mu}$ are massive by virtue of vortex excitations
being bound in finite loops.
The massive character of $v_\mu$ and $a_{\mu}$
protects the coherent TF excitations from being
smeared by vortex fluctuations. The coupling of TF's
to the gauge fields $v_\mu$ and $a_{\mu}$ is
{\em irrelevant}. This is our inverted Fermi liquid phase.

In the normal (pseudogap) state, the situation changes
dramatically. The bare propagators for $v_\mu$ and $a_{\mu}$
are now massless but the renormalization by the medium of
TF's {\em screens} these bare propagators
and still keeps $v_\mu$ massive. Thus,
TF coupling to the Doppler shift and ``spin-waves" remains
{\em irrelevant} even in the normal state. The Berry gauge field
$a_{\mu}$, however, is now truly {\em massless} since
the spin polarization in the medium of TF's
{\em cannot} fully screen the massless bare propagator. Instead, by
computing the TF polarization bubble, we find
${\cal D}_a^{-1}\propto {1\over 8}\sqrt{\omega^2 + q^2}$ for
$({\bf q},i\omega)\to 0$ \cite{gusynin}; stiffer than the
Maxwellian form (\ref{2d'},{\ref{2+1d}), but still massless.
The massless gauge field $a_{\mu}$ produces strong scattering at
low energies and affects qualitatively the spectral
properties of TF's.

The low energy quasiparticles are located at the four nodal
points of the $d_{xy}$ gap function:
$(\pm k_F,0)$ and $(0,\pm k_F)$, hereafter denoted as $(1,\bar{1})$ and
$(2,\bar{2})$ respectively.
Linearizing the fermionic spectrum in the proximity of these
nodes leads to the effective Lagrangian
\begin{eqnarray}
{\cal L}_D&=&
\sum_{\alpha=1,\bar{1}}\Psi_\alpha^{\dag}[D_{\tau}
- iv_FD_x\sigma_3- iv_\Delta D_y\sigma_1]
\Psi_\alpha
\label{qed3} \\
&+&\sum_{\alpha=2,\bar{2}}\Psi_\alpha^{\dag}[D_{\tau}
- iv_FD_y\sigma_3- iv_\Delta D_x\sigma_1]\Psi_\alpha
+{\cal L}_0[a_\mu],
\nonumber
\end{eqnarray}
\noindent
where $\Psi_\alpha^{\dag}$ is a two-component nodal spinor,
$\alpha$ is a node index,
$D_\mu = \partial_\mu +ia_\mu$, and ${\cal L}_0$ is
given by (\ref{2+1d}).
We have dropped the Doppler gauge field $v_\mu$ since
it is massive both below and above $T_{sc}$ and irrelevant for
our purposes.

The Lagrangian (\ref{qed3}) and the
physics it embodies are our main results. In the normal state
$a_{\mu}$ becomes massless and the problem of quasiparticle
interactions with vortex fluctuations takes the form
of topological fermions interacting with massless ``berryons",
i.e.\ quanta of the Berry gauge field $a_\mu$.
We recognize the above theory as equivalent (apart from the
intrinsic anisotropy) to the Euclidean quantum electrodynamics of
massless Dirac fermions in (2+1)-dimensions (QED$_3$).
The remarkable feature of QED$_3$ is that it
naturally generates a non-Fermi liquid
phenomenology for its fermionic excitations.
This property of QED$_3$ has led to
previous suggestions that it should be in some form relevant
to cuprate superconductivity \cite{aichinson,kim}.
However, the {\em physical content}
of QED$_3$ as an effective low energy theory in this Letter is
entirely different from those earlier works.

We now discuss
the low energy phenomenology governed by the TF propagator:
${ G}^{-1}_\alpha ({\bf k},\omega) =
{G}^{-1}_{\alpha0} ({\bf k},\omega) - \Sigma_\alpha ({\bf k},\omega)$,
where ${G}^{-1}_{\alpha0}$ is a free Dirac propagator at node $\alpha$.
We first consider the $T=0$
case in the isotropic limit $(v_F=v_\Delta)$ where explicit results are
readily obtained. In this case $G_{\alpha 0}
^{-1}=\omega -v_F k_x\sigma_3 - v_\Delta k_y\sigma_1$ and we find
\begin{equation}
\Sigma_\alpha = {8\over 3\pi^2 N}(-\omega +v_F k_x\sigma_3
+ v_\Delta k_y\sigma_1)\ln\bigl(\Lambda /p\bigr),
\label{green}
\end{equation}
with $p= (-\omega^2 + v_F^2k_x^2 +v_\Delta^2k_y^2)^{1/2}$,
$\Lambda$  a high energy cutoff, and $N=2$ is the number of pairs of nodes.  

The essential feature of the TF propagator is the singular
behavior of the self-energy $\Sigma_\alpha ({\bf k},\omega)$
which arises from the massless
nature of the dressed berryon propagator ${\cal D}_a({\bf q},\omega)$
and is logarithmic in the leading order. This result
can be formalized as the leading term in a large $N$ expansion.
Ultimately, the resummation of such an expansion \cite{aichinson}
yields a power law singularity $G_\alpha\propto p^{\eta-1}$,
with a small exponent $\eta = -8/3\pi^2N$ \cite{gusynin}. For our purposes,
having to deal with both the
anisotropy and finite $T$, the leading order
form (\ref{green}) is more convenient since it allows
for explicit computation of various quantities. Once
we move beyond the leading order, the vertex corrections
to $\Sigma_\alpha $ are necessary and the algebra
becomes impenetrable. Furthermore,
the available experiments are unlikely to
distinguish between $\eta = 0^+$ and a small finite exponent.

The singularity in $\Sigma_\alpha$
heralds the breakdown of the Fermi liquid behavior in the normal
state. To see this consider Eq.\ (\ref{green}) for $E_{\bf k}\equiv
(v_F^2k_x^2 + v_\Delta^2k_y^2)^{1/2}\ll |\omega |$. We find
$\Sigma_\alpha \propto -(8/3\pi^2 N)\omega
\ln\bigl(\Lambda /\sqrt{-\omega^2}\bigr)$.
The residue of the fermion pole vanishes as $\omega\to 0$,
$Z(\omega)\sim 1/\ln |\omega |$, while its width
goes as $\Sigma_\alpha'' = -(4/3\pi N)|\omega |$.
This behavior is reminiscent of the MFL expression for the
self-energy, assumed on phenomenological grounds
by Varma, Abrahams and collaborators \cite{varma}.
Note, however, that
our QED$_3$ TF propagator implied by Eq.\ (\ref{green}) remains
{\em qualitatively different} from the MFL {\em Ansatz} \cite{varma},
both by the fact that $\ln\bigl(\Lambda /p\bigr)$
is replaced by a weak power law (thus {\em algebraic} Fermi liquid) 
and by the momentum dependence of $\Sigma_\alpha ({\bf k},\omega)$.
As shown below it is this combined 
momentum-frequency dependence that provides 
a natural explanation for some of the remarkable
features of the fermionic spectral function
in cuprates observed in the ARPES experiments \cite{valla}. Also, we emphasize 
that our results apply to the pseudogap phase below $T^*$. The physics of 
the normal state at higher temperatures is beyond the scope of our present 
theory.

Inspection of Eq.\ (\ref{green})
reveals that $\Sigma_\alpha$ has imaginary part only inside the cone
defined by $\omega^2>v_F^2k_x^2 + v_\Delta^2k_y^2$; outside this cone
$\Sigma_\alpha''$ vanishes. This implies that TF spectral function plotted
as a function of momentum at fixed $\omega$ (MDC) will be very sharp close to
the Fermi surface, while the corresponding energy distribution curve (EDC)
will be broad. This is illustrated in Fig.\ \ref{fig2} where we plot
the spectral function $A({\bf k},\omega)=\pi^{-1}{\rm Im}[G_\alpha
({\bf k},\omega)]_{11}$ deduced from Eq.\ (\ref{green}).
\begin{figure}[t]
\includegraphics[width=8cm]{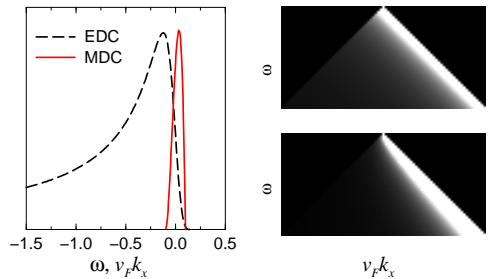}
\caption{\label{fig2}
Energy versus momentum distribution curves of $A({\bf k},\omega)$.
Left: EDC cut taken for ${\bf k}=0$ (coincident with a nodal point), 
and MDC cut taken for $\omega=0^{-}$ and $k_y=0$. 
Both curves have been broadened (by the 
same amount) to simulate the finite resolution of an ARPES experiment. Right:
The corresponding spectral function density plots for isotropic (top) and
anisotropic $v_F/v_\Delta=17$ (bottom) cases; to be compared with Figs. 1 and 
2 of Ref. \cite{valla}.}
\end{figure}
We note that precisely such
striking asymmetry between the EDC and MDC cuts is observed in the ARPES
data \cite{valla}.

These qualitative features of the spectral function survive at finite
temperature and away from the isotropic limit.
Unfortunately, away from this simple limit
the precise form of the TF propagator is not known:
as soon as the ``relativistic'' invariance of the $T =0$ problem
(\ref{qed3}) is lost, analytic calculations become intractable.
We find that for $T\ll \omega,E_{\bf k}$ the self energy retains its $T=0$
form Eq.\ (\ref{green}) with a small temperature correction. On the other hand
when $T\gg \omega,E_{\bf k}$ we find $\Sigma_\alpha'' \sim T$,
qualitatively consistent with the original MFL conjecture
$\Sigma_\alpha'' \sim {\rm max}(\omega,T)$. We note that such $T$--linear
scattering rate has been deduced from ARPES experiments \cite{valla}.

In ARPES one measures the spectral function of {\em real} electrons
not of TF's. While the inversion of the 
transformation (\ref{ft}) after the phases have been coarse-grained and 
replaced by the gauge fields is a daunting task, our theory insures 
the gauge invariance with respect to $a_\mu$ of the true electron propagator.
The simplest such gauge invariant propagator  
is $G_{11}^{\rm elec}(x,x') \approx \langle\exp (i\int^{x'}_x ds_\mu
a_\mu) [\tilde\Psi(x)\tilde\Psi^\dagger(x')]_{11}\rangle$,
where $x = ({\bf r},\tau)$.
By employing a gauge in which the line integral of $a_\mu$ vanishes
 \cite{rantner}, we have computed the asymptotic 
behavior of $G^{\rm elec}(x,x')$.
We find \cite{qed3long} that it exhibits a power law singularity with
the exponent  $\eta'=2\eta=-16/3\pi^2 N$. This strongly suggests 
that the true electron propagator, whose precise form within QED$_3$ is 
unknown at present, will exhibit a power law with small positive exponent.

In conclusion, we argued that the pseudogap regime in cuprates
can be modeled as a phase disordered $d$-wave superconductor.
Such assumption naturally leads to a QED$_3$ theory for the massless
Dirac ``topological'' fermions interacting with a massless
gauge field of vortex ``berryons''. Coupling to the massles gauge field
destroys the Fermi liquid pole in the fermion propagator and generates 
algebraic Fermi liquid. Lacking any energy or length scale this
theory can be thought of as being  {\em critical} independently
of the actual doping level $x$.
Below $T^*$ the low energy spectral properties of the fermions
are therefore regulated by a {\em quantum critical line}. In this regime
the low energy fermiology, including thermodynamics,
transport, and density and current responses are all controlled
by the universal properties of topological
fermions and vortex berryons encoded in the anisotropic QED$_3$
Lagrangian (\ref{qed3}).
Eventually, this peculiar quantum critical behavior gives way to
the actual superconducting phase at $T_{sc}(x)$ and the
Fermi liquid character of the nodal quasiparticles
is restored as vortices bind into finite loops.
At very low doping, hole Wigner crystal, SDW, and other low $T$ phases
become possible, reflecting the strong Mott-Hubbard correlations.

The authors are indebted to G. Baskaran, M. P. A. Fisher,
S. Girvin, I. Herbut,
C. Kallin, S. A. Kivelson, A. J. Millis, H. Monien, W. Rantner,
A. Sudb\o, A.-M. Tremblay, O. Vafek, P. B. Wiegmann and A. Zee for helpful
discussions. This research was supported in part by NSF Grants
PHY99-07949 (ITP) and DMR00-94981 (ZT).

\end{document}